\documentclass[runningheads]{llncs}
\usepackage[T1]{fontenc}
\usepackage{amsmath}
\usepackage{newtxtext}
\usepackage[varvw]{newtxmath}       % selects Times Roman as basic font

\usepackage[normalem]{ulem}
\usepackage{tcolorbox}
\usepackage{algorithm}
\usepackage[noend]{algpseudocode}
\usepackage{graphicx}
\usepackage{multirow}

\graphicspath{{../paper/}}

\begin{document}
\title{Quantum-Inspired Pairwise Ranking Aggregation via Active LLM Feedback}
\titlerunning{Quantum-Inspired Pairwise RA via Active LLM Feedback}
\author{Quan Mai\inst{1,2} \and Susan Gauch\inst{1} \and Gibson Nkhata\inst{1}}
\authorrunning{Q. Mai et al.}
\institute{University of Arkansas, Fayetteville, AR 72701, USA \and
Applied AI, Walmart Inc., Bentonville, AR 72716, USA}
\maketitle              % typeset the header of the contribution
\begin{abstract}
Ranking aggregation from multiple retrievers is a powerful yet under-explored technique in modern retrieval and RAG systems. We introduce LLM-IRA, the first extension of the quantum-inspired interactive ranking aggregation framework~\cite{hu2024qi} to document reranking. By leveraging large language models (LLMs) solely for pairwise judgments on high-disagreement document pairs, our method learns fusion weights in a principled, interactive way. The final ranking is a simple, ultra-fast pointwise linear fusion of scores, no full pairwise matrix or expensive listwise prompting is required. Experiments on the BEIR benchmark and TREC Deep Learning tracks with an open-weight instruction-tuned judge (Llama-3.1-8B-Instruct) show that LLM-IRA matches or numerically exceeds strong static fusion baselines on 5 of 6 evaluated benchmarks. We report two weight-update variants: momentum blending and a full accumulated-history refit, which together reach statistical significance (Wilcoxon signed-rank, $p<0.05$) over the best unsupervised aggregator on 4 of 6 benchmarks (DL19 under momentum; DL20, DBpedia, and NFCorpus under accumulated refit, which is the stronger of the two configurations overall). On the benchmarks we profiled (DL19, DL20), it already clears the best unsupervised baseline within as few as 5 LLM calls per query, well before exhausting its typical $\sim$25--30 call budget. LLM-IRA also achieves cost-effectiveness comparable to state-of-the-art zero-shot LLM rerankers at an order of magnitude fewer LLM calls, and requires no training data or model fine-tuning.

\end{abstract}
\section{Introduction}

Ranking Aggregation (RA) is a crucial technique in information retrieval, designed to fuse multiple search strategies to improve robustness and generalization, as no single base ranker performs optimally across all domains. Existing RA methods broadly divide into two categories: unsupervised techniques such as Reciprocal Rank Fusion (RRF~\cite{cormack2009reciprocal}) and CombSUM, which are simple and cost-effective but lack external guidance for optimization; and fully-supervised approaches, including learning-to-rank models (RankNet~\cite{burges2006learning}, LambdaMART~\cite{burges2010ranknet}) or zero-shot listwise LLM rerankers (RankGPT~\cite{sun2023chatgpt}, RankLLM~\cite{sharifymoghaddam2025rankllm}). While the latter deliver superior performance, they demand either extensive labeled data for training or prohibitive inference costs (hundreds of LLM calls per query) through massive prompting.
Recently, a Quantum-Inspired Interactive Ranking Aggregation (QI-IRA) \cite{hu2024qi} framework was proposed for image retrieval (person re-identification), demonstrating that RA can be modeled as fusing multiple ``incomplete observations" into a ``complete observation" using quantum theory. QI-IRA achieved state-of-the-art performance with remarkably low labeling costs (as few as two user feedbacks per query). However, QI-IRA's pointwise relevance feedback (e.g., ``Is this item relevant or irrelevant?") is inadequate for document reranking, where strict relative ordering is required.
In this paper, we propose LLM-IRA, a novel framework that extends the quantum RA analogy from point-wise absolute relevance to pairwise relative relevance. We model document pairs as entangled quantum systems and replace costly human feedback with an efficient interactive LLM loop that dynamically samples only the highest-disagreement pairs. This enables our method to learn optimal fusion weights for heterogeneous base rankers while retaining the simplicity and speed of linear pointwise fusion. We report two weight-update rules for turning LLM preferences into fusion weights: \textit{momentum blending} and a \textit{full accumulated-history refit}, and find the latter the stronger configuration overall. Experiments across two independent LLM judges demonstrate that LLM-IRA matches or outperforms static fusion baselines on the majority of tested benchmarks under either rule, and matches zero-shot LLM rerankers at a fraction of the cost.

The remainder of this paper is organized as follows: Section \ref{sec: related} reviews related work; Sections \ref{sec: recap} and \ref{sec: methodology} recap QI-IRA and detail our method, respectively; Sections \ref{sec: experiments} and \ref{sec: results} present experiments and results; and Section \ref{sec: conclusion} concludes.

\section{Related Work} \label{sec: related}
\noindent\textbf{Rank aggregation} RA combines outputs from multiple base rankers to improve overall performance. Unsupervised methods, such as RRF and CombSum/CombMNZ~\cite{fox1994combination}, are lightweight and widely used in hybrid retrieval pipelines, requiring no training or labels. These approaches fuse ranks or scores heuristically and perform well in diverse settings but lack adaptability to query-specific nuances. Supervised RA treats fusion as a learning-to-rank (LTR) problem. Classic models like RankNet~\cite{burges2006learning} use pairwise losses to learn preferences, while LambdaMART~\cite{burges2010ranknet} optimizes listwise metrics via gradient boosting. These methods achieve high accuracy with sufficient labeled data but incur heavy annotation and training costs, limiting their use in zero-shot or dynamic scenarios.

\noindent\textbf{LLM zero-shot document reranking} The rise of LLMs has produced a parallel and highly practical family of zero-shot rerankers. These fall into three main paradigms: pointwise scoring \cite{liang2022holistic,sachan2022improving}, pairwise comparison \cite{qin2024large}, and listwise reranking \cite{ma2023zero,pradeep2023rankvicuna,sun2023chatgpt}. While effective on benchmarks like BEIR and MS MARCO, these methods scale poorly: listwise prompting requires linear or quadratic LLM calls relative to candidate size, leading to high token costs. LLM-IRA uses LLMs more sparingly, only for targeted pairwise judgments on disagreed pairs, to guide fusion weights, achieving comparable performance at a fraction of the expense.

\noindent\textbf{Interactive relevance feedback} has evolved from classical Rocchio-style pseudo-relevance feedback \cite{rocchio1971relevance} and its neural variants \cite{li2018nprf} toward LLM-as-Judge techniques \cite{zheng2023judging}. Recent studies explore pairwise and listwise LLM judgments for query rewriting or preference alignment \cite{zhao2025permutative,liu2025lipo}, yet they typically treat the LLM as a full reranker or query expander rather than as a lightweight supervisor that learns fusion weights for cheaper base retrievers.

\noindent\textbf{LLM-guided rank fusion} Most closely related to our work, LLM-RankFusion~\cite{shen2024llmrankfusion} uses in-context learning with calibration to estimate pairwise preference probabilities and aggregate multiple rankers. While it achieves strong effectiveness, it requires comparing \emph{all} pairs in the candidate pool (quadratic LLM cost), making it impractical for large pools. In contrast, LLM-IRA queries the LLM only on the small subset of pairs where base rankers disagree, reducing cost by an order of magnitude while maintaining competitive effectiveness.

While unsupervised RA is efficient but static, supervised LTR and LLM rerankers are powerful but costly, and interactive feedback lacks principled fusion mechanisms. LLM-IRA bridges these by extending QI-IRA's quantum-inspired framework with low-cost LLM pairwise interactions, enabling adaptive aggregation without training or excessive prompting.

\section{Preliminaries: QI-IRA Recap} \label{sec: recap}
The original QI-IRA framework models each gallery item as a quantum state evaluated under observation operators corresponding to base rankers. Each basic ranking score \( s_m(d) \) is interpreted as the collapse probability into a relevant state. Gleason's theorem justifies that the optimal fused score is a linear combination:
\[
s^*(d) = \sum_{m=1}^M w_m \cdot s_m(d)
\]
Given positive (\(\Psi^+\)) and negative (\(\Psi^-\)) feedback items, weights are updated as:
\[
\bar{w}^m = \frac{1}{|\Psi^+|} \sum_{d_i \in \Psi^+} s_m(d_i) - \frac{1}{|\Psi^-|} \sum_{d_j \in \Psi^-} s_m(d_j)
\]
followed by momentum smoothing:
\[
w_t^m = \gamma \bar{w}^m + (1-\gamma) w_{t-1}^m
\]
where $\gamma$ regulates the degree of weight updating. This forms the foundation for our pairwise extension.

\section{The Pairwise Quantum Analogy}\label{sec: methodology}
QI-IRA focuses on pointwise absolute relevance with bases $|R\rangle$ (relevant) and $|\tilde{R}\rangle$ (irrelevant). For document reranking, which emphasizes strict relative ordering, we redefine the quantum model as follows:

\textit{The Particle System:} Each microscopic particle represents a document pair (Document A and Document B), with joint state vector $|d_{A,B}\rangle$.

\textit{The Observation Operators:} The base rankers (e.g., BM25, dense, sparse) serve as quantum observation operators.

\textit{The Orthogonal Bases:} For each ranker $m$, the 2-dimensional state space captures relative relevance via orthogonal bases: (1) $|A \succ_m B\rangle$ (Document A is strictly more relevant than B), and (2) $|B \succ_m A\rangle$ (Document B is strictly more relevant than A).

This shift enables modeling pairwise preferences while preserving the quantum-inspired framework.

\noindent\textbf{From 1D scores to pairwise collapse probabilities}
To map pointwise scores (e.g., BM25 scores) into the pairwise quantum state, we adapt the pairwise logit from RankNet-style LTR:

\noindent\textit{Pairwise collapse probability:} The probability that A wins over B under ranker $m$ is:
\[
s_{m(A,B)} = \frac{1}{1 + e^{-(score_A - score_B)}}
\]

The amplitudes sum to 1 in squared form, yielding:
$$
|d_{A,B}\rangle = \sqrt{s_{m(A,B)}} |A \succ_m B\rangle + \sqrt{1 - s_{m(A,B)}} |B \succ_m A\rangle$$

\noindent\textbf{Aggregation via Gleason's theorem}
QI-IRA uses Gleason's theorem to derive linear fusion of pointwise scores. This extends directly to our pairwise case: The density operator for ranker $m$ is \(\rho_m = |A \succ_m B\rangle\langle A \succ_m B|\). The aggregated operator is:
\[
\rho^* = \sum_{m=1}^M w_m \rho_m
\]
Substituting into Gleason's theorem yields the fused pairwise probability:
\[
s^*_{(A,B)} = \sum_{m=1}^M w_m s_{m(A,B)}
\]
Thus, the final ranking is derived from the aggregated pairwise probabilities, maintaining theoretical optimality. A detailed mathematical derivation extending Gleason's theorem to our pairwise quantum observation framework is provided in the Appendix~\ref{sec:appendix}.

\noindent\textbf{Interactive LLM feedback loop}
Starting with uniform weights, we compute an initial fused ranking. From the top-$r$ documents, we sample $K$ pairs with high ranker disagreement (e.g., where BM25 prefers A over B, but dense prefers B over A). These pairs are queried to the LLM: ``Given the query, which document is more relevant: A or B?'' The LLM's judgment (e.g., W $\succ$ L, where W is the winner and L the loser) acts as a definitive measurement, collapsing the pair's state.

\noindent\textbf{Weight adjustment}
Only LLM judgments with confidence (logit gap) at or above a threshold $\tau$ are retained in the valid preference set $V$; low-confidence judgments are discarded as noise rather than assumed to indicate an explicit ``both irrelevant'' outcome (see Limitations). For each retained judgment (W $\succ$ L), define the positive score \(s_m^+\) as the probability ranker $m$ correctly predicted the order (\(s_{m(W,L)}\)) and the negative score \(s_m^-\) as the opposite (\(s_{m(L,W)}\)). For a batch of $|V|$ pairs, the provisional weight is:
$$\bar{w}_m^t = \frac{1}{|V|} \sum_{k=1}^{|V|} (s_{m(W_k, L_k)} - s_{m(L_k, W_k)})$$
Because $s_{m(L,W)} = 1 - s_{m(W,L)}$, this simplifies to:
$$\bar{w}_m^t = \frac{1}{|V|} \sum_{k=1}^{|V|} (2 \times s_{m(W_k, L_k)} - 1) \;\in [-1, 1]$$
This ensures that a ranker providing a highly confident, correct probability receives a massive reward, a highly confident incorrect probability receives a massive penalty, and an uncertain ranker receives no update. Because $\bar{w}_m^t$ can be negative for a systematically wrong ranker, we exponentiate and renormalize across rankers (softmax) before momentum blending, which guarantees $w_m^t > 0$ and $\sum_{m=1}^M w_m^t = 1$ at every round. We introduce a weight-sharpness parameter $\beta$ (default $\beta{=}1$) that scales the provisional weight before exponentiating:
$$\hat{w}_m^t = \frac{\exp(\beta \cdot \bar{w}_m^t)}{\sum_{m'=1}^M \exp(\beta \cdot \bar{w}_{m'}^t)}$$
At $\beta{=}1$, the softmax keeps weight spread across all rankers even under strong disagreement, which is a sensible default when base rankers are genuinely complementary. Raising $\beta$ lets the method commit more fully to a single dominant ranker when one consistently outperforms the others, at the cost of amplifying noise from small per-round preference batches.

Finally, we update the actual fusion weights using one of two rules. The default, \emph{momentum}, blends only the current round's softmax estimate with the previous weights, where $\gamma$ (default 0.7) controls the degree of updating to ensure stability:
$$w_m^t = \gamma \times \hat{w}_m^t + (1 - \gamma) \times w_m^{t-1}$$
The alternative, \emph{accumulated refit}, instead recomputes $\bar{w}_m^t$ (and hence $\hat{w}_m^t$) from \emph{every} confident preference observed so far rather than only the current round's batch, and adopts it directly as the new weight with no momentum blend: $w_m^t \leftarrow \hat{w}_m^t$. We report both variants in Table~\ref{tab:compare} ($\beta{=}1$/momentum vs.\ $\beta{=}3$/accumulated refit); under accumulated refit, $\gamma$ is unused. 
% Note that $\beta$ and the update rule are independent axes --- any combination is valid --- but we only report this one pairing of each, so the accumulated-refit row's gains over momentum in Table~\ref{tab:compare} cannot be cleanly attributed to the update rule alone versus the higher $\beta$ (see Limitations).

The loop repeats until the LLM budget is exhausted or no unsampled disagreeing pairs remain in the top-$r$. The entire process is shown in Algorithm~\ref{alg:llm_ira}.

\begin{algorithm}[t]
\caption{Pairwise LLM-IRA for Document Reranking}
\label{alg:llm_ira}
\textbf{Input:} Query $q$, documents $D$, base-ranker scores $\{score_m(d)\}_{m=1}^M$, budget $T$, zone size $top\_r$, pairs/round $K$, confidence threshold $\tau$, sharpness $\beta$, update rule $\in\{$momentum $(\gamma)$, accum.\ refit$\}$ \\
\textbf{Output:} Final aggregated ranking of $D$
\begin{algorithmic}[1]
\State Min-max normalize each ranker's scores; derive pairwise collapse probabilities $s_{m(A,B)}$ (\S\ref{sec: methodology}); init $w_m^0 \leftarrow 1/M$
\For{$t = 1, \dots, T$}
    \State Aggregate $s^*(d) \leftarrow \sum_m w_m^{t-1} score_m(d)$; isolate Top-$top\_r$ documents
    \State Select the $K$ pairs with highest cross-ranker disagreement in $s_{m(A,B)}$ within this pool
    \State Query the LLM on each pair; keep judgments with confidence gap $\geq \tau$ as valid set $V$
    \If{no unsampled disagreeing pairs remain in Top-$top\_r$}
        \State \textbf{break} \Comment{sampling exhausted}
    \ElsIf{$V = \emptyset$}
        \State \textbf{continue} \Comment{all pairs below $\tau$; resample next round}
    \EndIf
    \State Compute provisional weights $\bar{w}_m^t$ from $V$, softmax-normalize with sharpness $\beta$, and apply momentum ($\gamma$) or accumulated refit to obtain $w_m^t$ (\S\ref{sec: methodology})
    \If{$w_m^t$ converged vs.\ $w_m^{t-1}$}
        \State \textbf{break}
    \EndIf
\EndFor
\State \Return $D$ sorted by $s^*(d) \leftarrow \sum_m w_m^t score_m(d)$, descending
\end{algorithmic}
\end{algorithm}

\section{Experiment Setup} \label{sec: experiments}
\paragraph{Datasets}We evaluate our method on three widely-used benchmarks: TREC Deep Learning 2019/2020 \cite{craswell2019overview,craswell2020overview}, and BEIR benchmark~\cite{thakur2021beir}.

\noindent\textbf{Base retrievers and candidate pool} We employ 3 heterogeneous base retrievers ($M=3$): BM25~\cite{robertson2009probabilistic} (lexical baseline), a dense retriever, and SPLADE~\cite{formal2021splade} (learned sparse retrieval). BM25 scores are represented via a linear rank-based proxy ($score = N - rank$, where $N$ is the candidate pool size), matching the transform used for our unsupervised baselines, since raw BM25 scores are not retained in our retrieval cache; the dense and SPLADE rankers use their native similarity/dot-product scores. For our \emph{primary} experiments we use \textbf{nomic-embed-text-v1.5}~\cite{nussbaum2024nomic} as the dense ranker; we additionally report results with all-MiniLM-L6-v2~\cite{reimers2019sentence} and \textbf{BGE-large-en-v1.5}~\cite{bge2024} for retriever-quality robustness (Table~\ref{tab:robustness}), each applied with their prescribed query and document prefixes.

All methods operate on the same top-100 candidate pool retrieved by BM25, isolating the reranking/fusion effect from upstream retrieval quality.

\noindent\textbf{LLM judge and evaluation metrics} For the primary evaluation we use two open-weight, instruction-tuned decoder-only judges: \textbf{Llama-3.1-8B-Instruct}~\cite{dubey2024llama3} and \textbf{Qwen2.5-14B-Instruct}~\cite{qwen2025technical}, using each model's chat template and extracting next-token logits for labels ``A'' and ``B'' to preserve the pairwise probability framework. We additionally report earlier exploratory results with the \textbf{Flan-T5} family~\cite{longpre2023flan} (Large/XL/XXL variants) at the MiniLM retriever tier. The LLM budget is 100 pairwise queries per query (in practice $\sim$30 are used). The exact prompt is in Appendix~\ref{appendix: prompt}.
Re-ranking depth is $k=10$; the primary metric is NDCG@10 (reported as percentages).

\noindent\textbf{Baselines} We compare against three categories of baselines: (1) \textit{Unsupervised fusion}: the Comb* family~\cite{fox1994combination} (CombMIN, CombMAX, CombSUM, CombANZ, CombMNZ), BordaCount~\cite{aslam2001models}, and RRF~\cite{cormack2009reciprocal}; and (2) \textit{LTR with LLM supervision} (LTR-Logistic): logistic regression over the same LLM pairwise labels collected by our disagreement-based sampling, to isolate whether the quantum-inspired weight update adds value beyond standard LTR on the same supervision signal.

\begin{table}[]
\caption{NDCG@10 comparison across TREC DL and BEIR benchmarks, using \textbf{nomic-embed-text-v1.5} as the dense retriever and BM25+Dense+SPLADE ($M{=}3$) as base rankers throughout. LLM-IRA (Llama-3.1-8B) values use the final adopted configuration: contentious-zone $r{=}40$, $K{=}3$ pairs/round, and confidence-threshold $\tau{=}2.5$ on the LLM logit gap. Best overall in \textbf{bold}; best unsupervised \uline{\textit{underlined}} (ties shown on all tied cells). $^{*}$Statistically significant improvement over the best unsupervised aggregator in that column (Wilcoxon signed-rank test, paired per-query NDCG@10, $p<0.05$); see \S\ref{sec: results} for the full per-column significance breakdown. $^\ddagger$The accumulated-refit row uses weight-sharpness $\beta{=}3$ and the accumulated-refit update rule described in \S\ref{sec: methodology} (vs.\ $\beta{=}1$/momentum for the row above). Its $^{*}$ markers follow the same convention as the momentum row; see \S\ref{sec: results} for the full significance breakdown, including comparisons against the momentum row and LTR-Logistic.}
\label{tab:compare}
\centering
\small
\resizebox{\textwidth}{!}{%
\begin{tabular}{llcccccc}
\hline
\textbf{Category} & \textbf{Method} & \textbf{DL 19}  & \textbf{DL 20}  & \textbf{Covid}  & \textbf{SciFact} & \textbf{DBpedia} & \textbf{NFCorpus} \\ \hline
\multirow{7}{*}{\rotatebox{90}{\scriptsize Unsupervised}}
& CombMIN       & 67.51                 & 63.08                 & 76.99                 & 71.53                 & 41.14                 & 37.23                 \\
& CombMAX       & 60.04                 & 55.89                 & 68.34                 & 69.65                 & 37.02                 & 34.69                 \\
& CombSUM       & 69.56                 & \uline{\textit{65.24}} & 78.00                 & \uline{\textit{73.26}} & \uline{\textit{43.31}} & 37.64                 \\
& CombANZ       & \uline{\textit{69.62}} & 65.19                 & 78.00                 & 73.21                 & 43.24                 & 37.55                 \\
& CombMNZ       & 69.53                 & \uline{\textit{65.24}} & 78.00                 & \uline{\textit{73.26}} & \uline{\textit{43.31}} & \uline{\textit{37.70}} \\
& BordaCount    & 67.94                 & 63.18                 & 79.27                 & 72.38                 & 41.69                 & 36.92                 \\
& RRF           & 68.80                 & 64.56                 & \uline{\textit{80.07}} & 72.90          & 42.44                 & 36.98                 \\ \hline
\multirow{1}{*}{\rotatebox{90}{\scriptsize LTR}}
& LTR-Logistic   & \textbf{70.98}  & 67.02  & \textbf{82.02}  & 72.63  & 43.22  & 36.69  \\ \hline
\multirow{2}{*}{\rotatebox{90}{\scriptsize Ours}}
& LLM-IRA (Llama-3.1-8B, momentum) & 70.42$^{*}$ & 65.95 & 80.04 & 73.47 & 43.66 & 37.81 \\
& LLM-IRA (Llama-3.1-8B, $\beta{=}3$, accum.\ refit)$^\ddagger$ & 70.45 & \textbf{67.67}$^{*}$ & 80.57 & \textbf{73.70} & \textbf{44.29}$^{*}$ & \textbf{38.01}$^{*}$ \\ \hline
\end{tabular}%
}
\end{table}

\noindent\textbf{Parameters} We set the top-$r$ parameter (contentious zone) to 40 for selecting disagreement pairs, choose the $K{=}3$ highest-disagreement pairs to query the LLM in each round, and only fold a pairwise judgment into the weight update if the judge's logit gap exceeds a confidence threshold $\tau{=}2.5$. We arrived at this configuration by sweeping $r$ from 10 to 40: small $r$ starves the sampler under a fixed per-query LLM budget by repeatedly re-querying an already-resolved contentious zone, while $r{=}40$ consistently gives the sampler enough candidate pairs to keep surfacing genuinely high-disagreement pairs as rounds proceed. $\tau{=}2.5$ was found to weakly dominate the laxer $\tau{=}2.0$ default (never worse, occasionally marginally better) across datasets, and is adopted as the final setting; $r$ and $K$ must remain small enough to stay computationally cheap, as sampling $K$ pairs out of the zone requires $O(K \log r^2)$ time complexity before feeding them to the LLM.

All experiments were conducted in a single run on 2 NVIDIA Tesla T4 GPUs, 16GB of VRAM each. Code will be made available upon acceptance.

\begin{table}[]
\caption{Cost--effectiveness comparison on TREC DL 2019, at the MiniLM/Flan-T5-Large tier (matching the LLM reranker baselines' judge). $^*$LLM reranker numbers reproduced from Wang et al.~\cite{wang2025realm} using Flan-T5-Large. The LLM-IRA row uses the accumulated-refit update rule ($\beta{=}3$) with BM25+Dense (MiniLM) base rankers ($M{=}2$), contentious-zone $r{=}20$, and the default confidence threshold ($\tau{=}2.0$); this secondary tier is retained for a judge-matched cost comparison, distinct from the primary nomic/Llama-3.1-8B tier in Table~\ref{tab:compare}. LTR-Logistic is omitted here: it uses the identical disagreement-based sampling and LLM-call budget as LLM-IRA ($\sim$27 calls/query), so its cost profile is the same; see Table~\ref{tab:compare} for its NDCG@10 comparison.}
\label{tab:llm-rerankers}
\centering
\small
\begin{tabular}{llll}
\hline
TREC DL 2019              & \textbf{NDCG@10} & \textbf{\# Calls} & \textbf{\#Tokens} \\ \hline
\multicolumn{4}{l}{\textit{LLM Rerankers (Flan-T5-Large)$^*$}} \\
TourRank         & 48.2             & 130.0                 & 92571                               \\
PRP-Graph        & 65.8             & 482.7                 & 221782                              \\
Setwise-HeapSort & 66.9             & 125.3                 & 40450                               \\ \hline
\multicolumn{4}{l}{\textit{LLM-guided Fusion (Flan-T5-Large)}} \\
% LLM-RankFusion   & 71.14            & 380.0                 & 85158                               \\
\textbf{LLM-IRA (Ours)}  & 66.43   & 26.95                 & 6003                                \\ \hline
\end{tabular}
\end{table}

\section{Results} \label{sec: results}
\begin{figure}[t]
  \centering
  \includegraphics[width=0.8\textwidth]{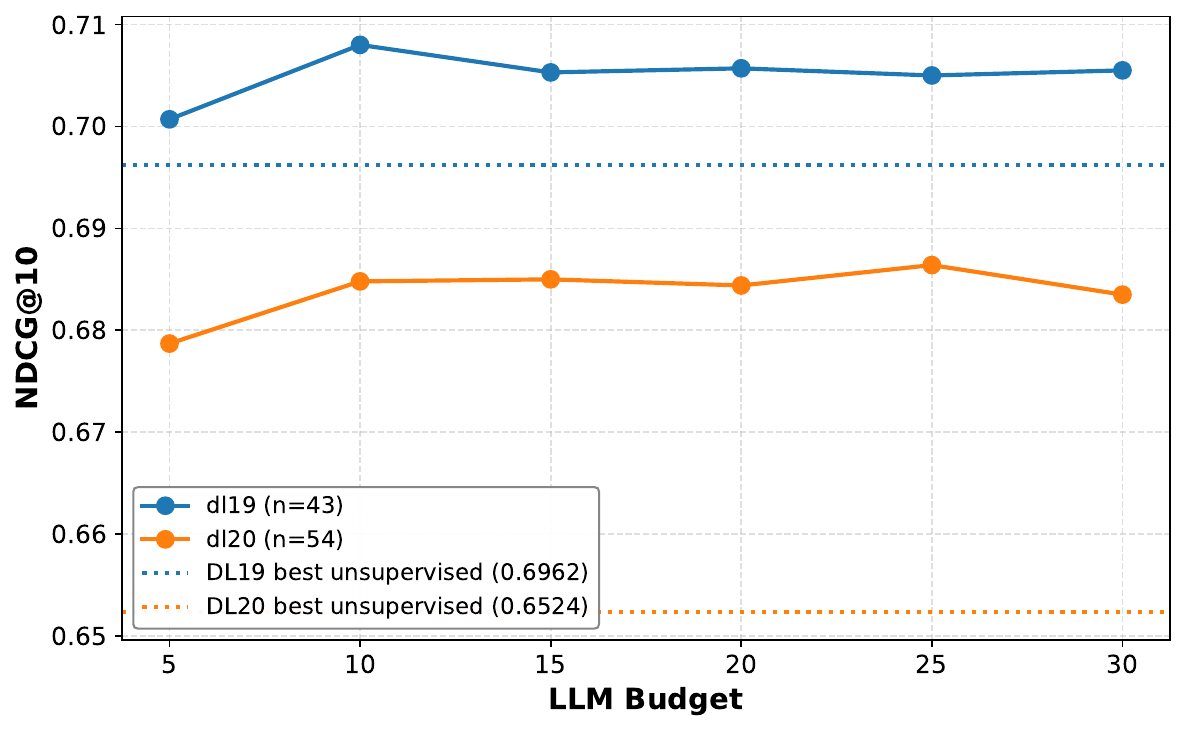}
  \caption{NDCG@10 vs.\ LLM-call budget (5--30 calls) at the primary nomic/Llama-3.1-8B-Instruct tier, under the accumulated-refit ($\beta{=}3$) configuration, for DL19 ($n{=}43$) and DL20 ($n{=}54$); checkpoints are read from per-round logs within a single full-budget run. Dotted lines mark each column's best unsupervised aggregator from Table~\ref{tab:compare} (CombANZ 69.62 for DL19, CombSUM/CombMNZ 65.24 for DL20).}
  \label{fig:budget}
\end{figure}

Unless noted otherwise, results below use the nomic/Llama-3.1-8B-Instruct tier of Table~\ref{tab:compare}, which we treat as canonical (Figure~\ref{fig:budget} included). Table~\ref{tab:llm-rerankers} reports cost-profile results at the MiniLM/Flan-T5-Large tier (matching the reranker baselines' judge), and Table~\ref{tab:robustness} reports additional retriever- and judge-scale robustness checks (MiniLM/BGE-large tiers).
\paragraph{Comparison with Unsupervised RA}
\textbf{LLM-IRA (momentum) matches or beats the best unsupervised aggregator on 5 of 6 benchmarks, but the edge is statistically significant only on DL19.} At the nomic/Llama-3.1-8B tier, using $\sim$30 LLM calls per query, it trails only on Covid (by 0.03 NDCG@10). Against each column's best unsupervised aggregator (Wilcoxon, paired per-query NDCG@10), the win on DL19 (vs.\ CombANZ, $p{=}0.041$, $\Delta{=}{+}0.0080$, $n{=}43$) is borderline given the sample size; elsewhere the gap is not significant ($p$ from 0.052 on DBpedia to 0.94 on Covid).

\textbf{LLM-IRA reliably beats the weak unsupervised aggregators, but is only sometimes ahead of the strong ones.} It significantly beats CombMIN and CombMAX on all six benchmarks ($p<0.01$), and BordaCount/RRF on four of six (not Covid or SciFact). Against the strongest aggregators (CombSUM/CombANZ/CombMNZ, themselves near-ties with each other): significant on DL19 ($p{=}0.016$--$0.041$) and Covid ($p<0.01$ against all three); not significant on DL20 or SciFact; on DBpedia/NFCorpus only CombANZ clears significance ($p{=}0.025$, $0.0099$) while CombSUM/CombMNZ narrowly miss ($p{=}0.052$, $0.058$/$0.19$).

\textbf{A second configuration --- accumulated-history weight refit at $\beta{=}3$ --- beats momentum on all six benchmarks and is the strongest configuration overall on three of them.} Gains over momentum range +0.03 to +1.72 NDCG@10, significant on DL20 ($p{=}0.0031$) and DBpedia ($p{=}0.0225$); not on DL19, Covid, or SciFact ($p{=}0.86$, $0.35$, $0.51$). Against the best unsupervised aggregator it is significant on DL20 ($p{=}0.011$), DBpedia ($p{=}0.0056$), and NFCorpus ($p{=}0.019$) --- marked $^{*}$ in Table~\ref{tab:compare} --- not on DL19, Covid, or SciFact ($p{=}0.16$, $0.36$, $0.20$). Against LTR-Logistic it leads on four of six (DL20, SciFact, DBpedia, NFCorpus; significant on the latter two, $p{=}0.02$ and $p{<}0.001$) and trails on DL19 and Covid (neither significant, $p{=}0.29$ and $0.078$; Covid is the largest numerical deficit, 80.57 vs.\ 82.02). It is never significantly worse than momentum, LTR-Logistic, or the best unsupervised aggregator on any benchmark --- a reasonable case for adopting it going forward.

\paragraph{Impact of LLM budget}
\textbf{LLM-IRA exceeds the best unsupervised aggregator within just 5 LLM calls per query.} Figure~\ref{fig:budget} reports this at the primary nomic/Llama-3.1-8B-Instruct, accumulated-refit tier. On DL19 (best unsupervised: CombANZ, 69.62), LLM-IRA reaches 70.07 at a budget of only 5 calls and is essentially flat thereafter (70.50--70.80 across 10--30 calls). On DL20 (best unsupervised: CombSUM/CombMNZ, 65.24), LLM-IRA reaches 67.87 at 5 calls and drifts within 68.35--68.64 across 10--30 calls. Both curves clear the best unsupervised aggregator at the smallest budget tested and plateau rather than peaking at a specific call count, indicating most of the gain is captured from the first handful of high-disagreement pairs.
% \paragraph{Comparison with LLM-guided fusion}
% \textbf{LLM-IRA matches LLM-RankFusion's signal at a fraction of the LLM-call cost.} LLM-RankFusion~\cite{shen2024llmrankfusion} builds a complete pairwise preference matrix over the top-$r$ candidates using calibrated LLM queries (both orderings per pair); with $r=20$ this requires $2\times\binom{20}{2}=380$ LLM calls per query --- over 14$\times$ more than LLM-IRA's $\sim$27 calls. We also compare against LTR-Logistic, a logistic regression baseline that receives the \emph{same} LLM pairwise labels from our disagreement-based sampling and fits a single fusion weight via maximum likelihood, isolating the contribution of the quantum-inspired weight update from the sampling strategy. Its LLM-call cost is identical to LLM-IRA's (same sampling strategy and budget, $\sim$27 calls/query), so we omit it from the cost comparison (Table~\ref{tab:llm-rerankers}) and compare it on NDCG@10 only, in Table~\ref{tab:compare}.

\paragraph{Cost against LLM rerankers}
\textbf{LLM-IRA exceeds PRP-Graph's effectiveness at 5.6\% of its LLM calls, and comes within 0.5 NDCG@10 points of the SOTA HeapSort reranker at about a fifth of its cost.} We evaluate against computationally expensive LLM rerankers --- listwise TourRank~\cite{chen2025tourrank}, pairwise PRP-Graph~\cite{luo2024prp}, and setwise HeapSort~\cite{zhuang2024setwise}, reproduced from Wang et al.~\cite{wang2025realm} using Flan-T5-Large; our own LLM-IRA row (accumulated refit, $\beta{=}3$) uses the same MiniLM/Flan-T5-Large tier for a judge-matched comparison. As shown in Table~\ref{tab:llm-rerankers}, LLM-IRA (66.43 NDCG@10) exceeds PRP-Graph (65.8) using only 5.6\% of its LLM calls (27 vs.\ 483) and 2.7\% of its prompt tokens, and trails HeapSort (66.9) by just 0.47 NDCG@10 points using about a fifth of the LLM calls (27 vs.\ 125) and 14.8\% of the prompt tokens.
\paragraph{Robustness to retriever and judge quality}
\textbf{LLM-IRA's gains hold across retriever tiers and judge scales, and it significantly beats RRF at the primary nomic tier on DL20 (numerically ahead on DL19 and Covid as well).} Table~\ref{tab:robustness} reports NDCG@10 across retriever tiers and judges, alongside RRF at each tier. Switching the dense retriever from MiniLM to BGE-large raises the RRF ceiling, as expected. At the BGE-large tier, LLM-IRA (Llama-3.1-8B) beats RRF by +0.28 (DL19) and +1.11 (DL20), trails by 0.17 (Covid); only DL20 is significant ($p{=}0.0497$, $n{=}54$; DL19 $p{=}0.62$, Covid $p{=}0.96$). Llama-3.1-8B vs.\ the larger Qwen2.5-14B at the same tier shows near-identical effectiveness (largest gap 0.65, DL20); judge scale beyond $\sim$8B is not the binding constraint. The nomic/Llama-3.1-8B row reuses the accumulated-refit ($\beta{=}3$) run from Table~\ref{tab:compare}: vs.\ RRF it is significant on DL20 ($p{=}0.0008$), not significant on DL19 ($p{=}0.0594$) or Covid ($p{=}0.3643$) though numerically ahead on both --- a different comparison than Table~\ref{tab:compare}'s marker, which is against each column's best unsupervised aggregator rather than RRF specifically. Swapping the judge down to the far smaller, encoder-decoder Flan-T5-Large (770M) or Flan-T5-XL (3B) at this same nomic tier gives NDCG@10 on par with (if anything, marginally above) Llama-3.1-8B-Instruct --- 70.58/68.04/80.68 and 70.64/68.38/80.19 (DL19/DL20/Covid) respectively, both numerically ahead of RRF on all three --- extending the judge-scale-insensitivity finding down roughly an order of magnitude in parameter count, though we did not run significance tests for these two rows.

\begin{table}[h]
\caption{Robustness of LLM-IRA across retriever and judge configurations ($M{=}3$: BM25, Dense, SPLADE). $^{*}$Significant vs.\ RRF (Wilcoxon $p{=}0.0497$); DL19 and Covid gaps for Llama are not significant ($p{=}0.62$, $p{=}0.96$). $^{\S}$Significant vs.\ RRF at the nomic tier (Wilcoxon $p{=}0.0008$ DL20, $n{=}54$); the DL19 ($p{=}0.0594$, $n{=}43$) and Covid ($p{=}0.3643$) gaps are not significant, though LLM-IRA is numerically ahead of RRF on both (DL19 +1.65, Covid +0.50 NDCG@10). The nomic/Llama-3.1-8B-Instruct row reuses the accumulated-refit ($\beta{=}3$) run reported as LLM-IRA in Table~\ref{tab:compare}.}
\label{tab:robustness}
\centering
\small
\begin{tabular}{llccc}
\hline
\textbf{Retriever} & \textbf{Judge} & \textbf{DL19} & \textbf{DL20} & \textbf{Covid} \\ \hline
\multicolumn{2}{l}{\textit{Best unsupervised (RRF)}} & & & \\
MiniLM   & --                    & 61.33 & 58.28 & 71.92 \\
BGE-large & --                   & 70.15 & 64.11 & 78.05 \\
nomic     & --                   & 68.80 & 64.56 & 80.07 \\ \hline
\multicolumn{2}{l}{\textit{LLM-IRA}} & & & \\
MiniLM    & Flan-T5-XXL & 63.15 & 62.58 & 74.23 \\
BGE-large & Llama-3.1-8B-Instruct & 70.43 & 65.22$^{*}$ & 77.88 \\
BGE-large & Qwen2.5-14B-Instruct  & 70.52 & 65.87 & 77.82 \\
nomic     & Llama-3.1-8B-Instruct (accum.\ refit) & 70.45 & 67.67$^{\S}$ & 80.57 \\
nomic     & Flan-T5-Large (accum.\ refit)         & 70.58 & 68.04 & 80.68 \\
nomic     & Flan-T5-XL (accum.\ refit)            & 70.64 & 68.38 & 80.19 \\ \hline
\end{tabular}
\end{table}

% \paragraph{Further Analysis}
% Across datasets, learned fusion weights show a clear domain effect: on TREC DL 2019 the dense retriever dominates heavily, whereas on BEIR Covid the margin shrinks substantially, as BM25 remains competitive on Covid's technical biomedical documents due to effective lexical matching. Weight evolution across interactive rounds converges within 7-10 rounds ($\sim$21-30 LLM calls); a full per-query weight-distribution and convergence-curve breakdown at the nomic/Llama-3.1-8B tier is left to an extended version of this work.

\section{Conclusion} \label{sec: conclusion}
We propose LLM-IRA, the first extension of quantum-inspired interactive ranking aggregation to document reranking. Under the Llama-3.1-8B-Instruct judge, LLM-IRA matches or numerically exceeds static rank aggregation methods on 5 of 6 evaluated benchmarks. Of the two weight-update rules we report, momentum reaches significance (Wilcoxon, $p<0.05$) over the best unsupervised aggregator on 1 of 6 benchmarks (DL19), while accumulated-history refit ($\beta{=}3$) extends this to 3 of 6 (DL20, DBpedia, NFCorpus), is never significantly worse than momentum on any benchmark, and is the configuration we recommend going forward. On the benchmarks we profiled (DL19, DL20), LLM-IRA already clears the best unsupervised baseline within as few as 5 LLM calls per query, well before exhausting its typical $\sim$25--30 call budget (Figure~\ref{fig:budget}). LLM-IRA enables efficient fusion of heterogeneous retrievers with minimal LLM feedback.

\section{Limitations}
While our framework shows strong potential, it presents several limitations for future research:

\paragraph{LLM noise} Even with capable 8B--14B instruction-tuned judges, pairwise preferences on contentious, high-disagreement document pairs remain noisy, and Covid shows this can narrow or flip the outcome against LLM-IRA relative to a strong static aggregator (the momentum configuration trails RRF there by 0.03 NDCG@10, Table~\ref{tab:compare}). Because the system directly translates the LLM's absolute judgment into mathematical weight adjustments, noisy or incorrect feedback can heavily penalize superior base rankers; a confidence-gated update (discarding low-margin judgments) mitigates but does not eliminate this risk, and the appropriate confidence threshold is judge-specific and currently set empirically rather than calibrated automatically.
\paragraph{Binary quantum state constraints} The current 2D quantum state space strictly models relative preference ($A \succ B$ or $B \succ A$). In reality, both documents might be entirely irrelevant to the query. Future work must mathematically integrate a ``both irrelevant" dimension into the model without violating the fundamental quantum rule that the sum of squared probability amplitudes must equal one.
\paragraph{Reversion to pointwise inference}  To maintain computational efficiency across large datasets, the final global inference step abandons the pairwise logic and reverts to standard 1D pointwise linear weighting. Future work could explore scalable methods for executing a strictly pairwise final aggregation.
% \paragraph{Confounded update-rule/sharpness comparison} The two configurations in Table~\ref{tab:compare} vary both the weight-update rule (momentum vs.\ accumulated refit) and the sharpness parameter $\beta$ (1 vs.\ 3) simultaneously, so the accumulated-refit row's gains cannot be attributed to the update rule alone versus the higher $\beta$. Isolating the two would require the missing $2{\times}2$ cells (momentum/$\beta{=}3$ and accumulated-refit/$\beta{=}1$), which we leave to future work.

\begin{credits}
\subsubsection{\discintname}
The authors have no competing interests to declare that are relevant to the content of this article.
\end{credits}

%
% ---- Bibliography ----
%
\bibliographystyle{splncs04}
\bibliography{custom}

@inproceedings{hu2024qi,
  title={QI-IRA: quantum-inspired interactive ranking aggregation for person re-identification},
  author={Hu, Chunyu and Zhang, Hong and Liang, Chao and Huang, Hao},
  booktitle={Proceedings of the AAAI conference on artificial intelligence},
  volume={38},
  number={3},
  pages={2202--2210},
  year={2024}
}

@inproceedings{sun2023chatgpt,
  title={Is ChatGPT good at search? investigating large language models as re-ranking agents},
  author={Sun, Weiwei and Yan, Lingyong and Ma, Xinyu and Wang, Shuaiqiang and Ren, Pengjie and Chen, Zhumin and Yin, Dawei and Ren, Zhaochun},
  booktitle={Proceedings of the 2023 conference on empirical methods in natural language processing},
  pages={14918--14937},
  year={2023}
}

@inproceedings{sharifymoghaddam2025rankllm,
  title={Rankllm: A python package for reranking with llms},
  author={Sharifymoghaddam, Sahel and Pradeep, Ronak and Slavescu, Andre and Nguyen, Ryan and Xu, Andrew and Chen, Zijian and Zhang, Yilin and Chen, Yidi and Xian, Jasper and Lin, Jimmy},
  booktitle={Proceedings of the 48th International ACM SIGIR Conference on Research and Development in Information Retrieval},
  pages={3681--3690},
  year={2025}
}

@article{burges2006learning,
  title={Learning to rank with nonsmooth cost functions},
  author={Burges, Christopher and Ragno, Robert and Le, Quoc},
  journal={Advances in neural information processing systems},
  volume={19},
  year={2006}
}

@article{burges2010ranknet,
  title={From ranknet to lambdarank to lambdamart: An overview},
  author={Burges, Christopher JC},
  journal={Learning},
  volume={11},
  number={23-581},
  pages={81},
  year={2010}
}

@inproceedings{cormack2009reciprocal,
  title={Reciprocal rank fusion outperforms condorcet and individual rank learning methods},
  author={Cormack, Gordon V and Clarke, Charles LA and Buettcher, Stefan},
  booktitle={Proceedings of the 32nd international ACM SIGIR conference on Research and development in information retrieval},
  pages={758--759},
  year={2009}
}

@article{fox1994combination,
  title={Combination of multiple searches},
  author={Fox, Edward A and Shaw, Joseph A},
  journal={NIST special publication SP},
  volume={243},
  year={1994},
  publisher={NATIONAL INSTIUTE OF STANDARDS \& TECHNOLOGY}
}

@article{craswell2020overview,
  author       = {Nick Craswell and
                  Bhaskar Mitra and
                  Emine Yilmaz and
                  Daniel Campos},
  title        = {Overview of the {TREC} 2020 deep learning track},
  journal      = {CoRR},
  volume       = {abs/2102.07662},
  year         = {2021},
  url          = {https://arxiv.org/abs/2102.07662},
  eprinttype    = {arXiv},
  eprint       = {2102.07662},
  bibsource    = {dblp computer science bibliography, https://dblp.org}
}

@inproceedings{craswell2019overview,
author = {Craswell, Nick and Mitra, Bhaskar and Yilmaz, Emine and Campos, Daniel and Voorhees, Ellen M.},
title = {Overview of the TREC 2019 deep learning track},
booktitle = {Text REtrieval Conference (TREC)},
year = {2020},
month = {March},
publisher = {TREC},
url = {https://www.microsoft.com/en-us/research/publication/overview-of-the-trec-2019-deep-learning-track/},
}

@article{thakur2021beir,
  title={Beir: A heterogenous benchmark for zero-shot evaluation of information retrieval models},
  author={Thakur, Nandan and Reimers, Nils and R{\"u}ckl{\'e}, Andreas and Srivastava, Abhishek and Gurevych, Iryna},
  journal={arXiv preprint arXiv:2104.08663},
  year={2021}
}

@inproceedings{zhuang2024setwise,
  title={A setwise approach for effective and highly efficient zero-shot ranking with large language models},
  author={Zhuang, Shengyao and Zhuang, Honglei and Koopman, Bevan and Zuccon, Guido},
  booktitle={Proceedings of the 47th International ACM SIGIR Conference on Research and Development in Information Retrieval},
  pages={38--47},
  year={2024}
}

@inproceedings{longpre2023flan,
  title={The flan collection: Designing data and methods for effective instruction tuning},
  author={Longpre, Shayne and Hou, Le and Vu, Tu and Webson, Albert and Chung, Hyung Won and Tay, Yi and Zhou, Denny and Le, Quoc V and Zoph, Barret and Wei, Jason and others},
  booktitle={International conference on machine learning},
  pages={22631--22648},
  year={2023},
  organization={PMLR}
}

@inproceedings{aslam2001models,
  title={Models for metasearch},
  author={Aslam, Javed A and Montague, Mark},
  booktitle={Proceedings of the 24th annual international ACM SIGIR conference on Research and development in information retrieval},
  pages={276--284},
  year={2001}
}

@article{liang2022holistic,
  title={Holistic evaluation of language models},
  author={Liang, Percy and Bommasani, Rishi and Lee, Tony and Tsipras, Dimitris and Soylu, Dilara and Yasunaga, Michihiro and Zhang, Yian and Narayanan, Deepak and Wu, Yuhuai and Kumar, Ananya and others},
  journal={arXiv preprint arXiv:2211.09110},
  year={2022}
}

@inproceedings{sachan2022improving,
  title={Improving passage retrieval with zero-shot question generation},
  author={Sachan, Devendra and Lewis, Mike and Joshi, Mandar and Aghajanyan, Armen and Yih, Wen-tau and Pineau, Joelle and Zettlemoyer, Luke},
  booktitle={Proceedings of the 2022 Conference on Empirical Methods in Natural Language Processing},
  pages={3781--3797},
  year={2022}
}

@article{ma2023zero,
  title={Zero-shot listwise document reranking with a large language model},
  author={Ma, Xueguang and Zhang, Xinyu and Pradeep, Ronak and Lin, Jimmy},
  journal={arXiv preprint arXiv:2305.02156},
  year={2023}
}

@article{pradeep2023rankvicuna,
  title={Rankvicuna: Zero-shot listwise document reranking with open-source large language models},
  author={Pradeep, Ronak and Sharifymoghaddam, Sahel and Lin, Jimmy},
  journal={arXiv preprint arXiv:2309.15088},
  year={2023}
}

@inproceedings{qin2024large,
  title={Large language models are effective text rankers with pairwise ranking prompting},
  author={Qin, Zhen and Jagerman, Rolf and Hui, Kai and Zhuang, Honglei and Wu, Junru and Yan, Le and Shen, Jiaming and Liu, Tianqi and Liu, Jialu and Metzler, Donald and others},
  booktitle={Findings of the Association for Computational Linguistics: NAACL 2024},
  pages={1504--1518},
  year={2024}
}

@article{rocchio1971relevance,
  title={Relevance feedback in information retrieval},
  author={Rocchio Jr, Joseph John},
  journal={The SMART retrieval system: experiments in automatic document processing},
  year={1971},
  publisher={Englewood Cliffs}
}

@inproceedings{li2018nprf,
  title={NPRF: A neural pseudo relevance feedback framework for ad-hoc information retrieval},
  author={Li, Canjia and Sun, Yingfei and He, Ben and Wang, Le and Hui, Kai and Yates, Andrew and Sun, Le and Xu, Jungang},
  booktitle={Proceedings of the 2018 Conference on Empirical Methods in Natural Language Processing},
  pages={4482--4491},
  year={2018}
}

@article{zheng2023judging,
  title={Judging llm-as-a-judge with mt-bench and chatbot arena},
  author={Zheng, Lianmin and Chiang, Wei-Lin and Sheng, Ying and Zhuang, Siyuan and Wu, Zhanghao and Zhuang, Yonghao and Lin, Zi and Li, Zhuohan and Li, Dacheng and Xing, Eric and others},
  journal={Advances in neural information processing systems},
  volume={36},
  pages={46595--46623},
  year={2023}
}

@inproceedings{zhao2025permutative,
  title={Permutative preference alignment from listwise ranking of human judgments},
  author={Zhao, Yang and Wang, Yixin and Yin, Mingzhang},
  booktitle={Proceedings of the 2025 Conference on Empirical Methods in Natural Language Processing},
  pages={310--334},
  year={2025}
}

@inproceedings{liu2025lipo,
  title={Lipo: Listwise preference optimization through learning-to-rank},
  author={Liu, Tianqi and Qin, Zhen and Wu, Junru and Shen, Jiaming and Khalman, Misha and Joshi, Rishabh and Zhao, Yao and Saleh, Mohammad and Baumgartner, Simon and Liu, Jialu and others},
  booktitle={Proceedings of the 2025 Conference of the Nations of the Americas Chapter of the Association for Computational Linguistics: Human Language Technologies (Volume 1: Long Papers)},
  pages={2404--2420},
  year={2025}
}

@inproceedings{reimers2019sentence,
  title={Sentence-bert: Sentence embeddings using siamese bert-networks},
  author={Reimers, Nils and Gurevych, Iryna},
  booktitle={Proceedings of the 2019 conference on empirical methods in natural language processing and the 9th international joint conference on natural language processing (EMNLP-IJCNLP)},
  pages={3982--3992},
  year={2019}
}

@book{robertson2009probabilistic,
  title={The probabilistic relevance framework: BM25 and beyond},
  author={Robertson, Stephen and Zaragoza, Hugo},
  volume={4},
  year={2009},
  publisher={Now Publishers Inc}
}

@inproceedings{chen2025tourrank,
  title={Tourrank: Utilizing large language models for documents ranking with a tournament-inspired strategy},
  author={Chen, Yiqun and Liu, Qi and Zhang, Yi and Sun, Weiwei and Ma, Xinyu and Yang, Wei and Shi, Daiting and Mao, Jiaxin and Yin, Dawei},
  booktitle={Proceedings of the ACM on Web Conference 2025},
  pages={1638--1652},
  year={2025}
}

@inproceedings{luo2024prp,
  title={Prp-graph: Pairwise ranking prompting to llms with graph aggregation for effective text re-ranking},
  author={Luo, Jian and Chen, Xuanang and He, Ben and Sun, Le},
  booktitle={Proceedings of the 62nd Annual Meeting of the Association for Computational Linguistics (Volume 1: Long Papers)},
  pages={5766--5776},
  year={2024}
}

@article{shen2024llmrankfusion,
  title={LLM-RankFusion: Mitigating Intrinsic Inconsistency in LLM-based Ranking},
  author={Shen, Yifan and Gao, Yingrui and Shao, Tao and Guo, Jiashu},
  journal={arXiv preprint arXiv:2406.00231},
  year={2024}
}

@misc{bge2024,
  title={{BGE M3-Embedding}: Multi-Lingual, Multi-Functionality, Multi-Granularity Text Embeddings Through Self-Knowledge Distillation},
  author={Chen, Jianlyu and Xiao, Shitao and Zhang, Peitian and Luo, Kun and Lian, Defu and Liu, Zheng},
  year={2024},
  eprint={2402.03216},
  archivePrefix={arXiv}
}

@article{nussbaum2024nomic,
  title={Nomic Embed: Training a Reproducible Long Context Text Embedder},
  author={Nussbaum, Zach and Morris, John X. and Duderstadt, Brandon and Mulyar, Andriy},
  journal={arXiv preprint arXiv:2402.01613},
  year={2024}
}

@article{dubey2024llama3,
  title={The {Llama} 3 Herd of Models},
  author={Dubey, Abhimanyu and others},
  journal={arXiv preprint arXiv:2407.21783},
  year={2024}
}

@inproceedings{wang2025realm,
  title={REALM: Recursive Relevance Modeling for LLM-based Document Re-Ranking},
  author={Wang, Pinhuan and Xia, Zhiqiu and Liao, Chunhua and Wang, Feiyi and Liu, Hang},
  booktitle={Proceedings of the 2025 Conference on Empirical Methods in Natural Language Processing},
  pages={23875--23889},
  year={2025}
}

@inproceedings{formal2021splade,
  title={{SPLADE}: Sparse Lexical and Expansion Model for First Stage Ranking},
  author={Formal, Thibault and Piwowarski, Benjamin and Clinchant, St{\'e}phane},
  booktitle={Proceedings of the 44th International ACM SIGIR Conference on Research and Development in Information Retrieval},
  pages={2288--2292},
  year={2021}
}

@article{qwen2025technical,
  title={Qwen2.5 Technical Report},
  author={Qwen Team},
  journal={arXiv preprint arXiv:2412.15115},
  year={2024}
}

\appendix

\section{Proof of Pairwise RA via Gleason's theorem}
\label{sec:appendix}

This appendix provides a rigorous mathematical derivation extending Gleason's theorem to pairwise relevance aggregation in document reranking. We model document pairs as quantum systems and show that the optimal fused pairwise probability is a linear weighted sum of individual ranker probabilities. The derivation assumes familiarity with basic quantum mechanics (state vectors, density operators, and trace operations) and Gleason's theorem, which states that any quantum probability measure on a Hilbert space can be expressed as the trace of the state operator with a density operator: \( p(E) = \operatorname{tr}[D \rho] \), where \( D \) is the system state operator and \( \rho \) is the measurement density operator for event \( E \).

\paragraph{Pairwise particle and state operator}
Consider a document pair (Document A and Document B) as the quantum system of interest. The joint state of the pair is represented by the state vector \( |d_{A,B}\rangle \). The corresponding density operator (state operator) is \( D_{A,B} = |d_{A,B}\rangle \langle d_{A,B}| \). This formulation treats the pair as a single entangled system in a 2-dimensional Hilbert space for binary relative relevance.

\paragraph{Pairwise observation bases}
Each base ranker (e.g., BM25, dense retriever, sparse retriever) acts as a measurement operator on the pair. For the \( m \)-th ranker, the measurement space is spanned by two orthogonal bases representing relative relevance
\begin{enumerate}
    \item \( |A \succ_m B\rangle \): The state where Document A is strictly more relevant than Document B according to ranker \( m \).
    \item \( |B \succ_m A\rangle \): The state where Document B is strictly more relevant than Document A according to ranker \( m \).
\end{enumerate}

These bases are orthonormal: \( \langle A \succ_m B | B \succ_m A \rangle = 0 \) and \( \langle A \succ_m B | A \succ_m B \rangle = \langle B \succ_m A | B \succ_m A \rangle = 1 \).

\paragraph{The Pairwise collapse probability}
The pairwise score \( s_{m(A,B)} \) from the \( m \)-th ranker is the probability that the system collapses into the state where A is more relevant than B:

\begin{equation}
s_{m(A,B)} = |\langle A \succ_m B | d_{A,B} \rangle|^2
\end{equation}

By Gleason's theorem, this probability can be reformulated using the density operator for the measurement:

\begin{enumerate}
    \item Define the density operator for the ``Document A is more relevant than B" event under ranker \( m \) as
    \begin{equation}
    \label{eq: x1}
        \rho_m = |A \succ_m B\rangle \langle A \succ_m B|
    \end{equation}
    \item Then,
  \begin{equation}
  \label{eq: x2}
  s_{m(A,B)} = \operatorname{tr}[D_{A,B} \rho_m]
  \end{equation}
\end{enumerate}

This holds because Gleason's theorem guarantees that all quantum probabilities are trace-form measures.

\paragraph{Aggregating pairwise rankings}
To fuse \( M \) rankers, we combine their ``incomplete" pairwise observations into a complete observation. Define the aggregated density operator as a linear combination of the individual density operators:

\begin{equation}
\rho^* = \sum_{m=1}^M w_m \rho_m = \sum_{m=1}^M w_m |A \succ_m B\rangle \langle A \succ_m B|
\end{equation}

where \( w_m \) are the fusion weights. For \( \rho^* \) to be a valid density operator, the \( w_m \) must form a probability distribution: \( w_m \geq 0 \) and \( \sum w_m = 1 \). As described in the main text (Section~\ref{sec: methodology}), the raw per-ranker statistic that drives \( w_m \) can be negative for a systematically wrong ranker, so we exponentiate and renormalize (softmax) before substituting into \( \rho^* \), guaranteeing this constraint holds at every round.

Applying Gleason's theorem to the aggregated operator gives the fused pairwise probability:

\begin{equation}
s^*_{(A,B)} = \operatorname{tr}[D_{A,B} \rho^*].
\end{equation}

Substituting the expression for \( \rho^* \):

\begin{equation}
\begin{split}
s^*_{(A,B)} &= \operatorname{tr}\left[ D_{A,B} \sum_{m=1}^M w_m |A \succ_m B\rangle \langle A \succ_m B| \right] \\ &= \sum_{m=1}^M w_m \operatorname{tr}[D_{A,B} |A \succ_m B\rangle \langle A \succ_m B| ].
\end{split}
\end{equation}

From Equations~\ref{eq: x1} and~\ref{eq: x2}, \( \operatorname{tr}[D_{A,B} |A \succ_m B\rangle \langle A \succ_m B| ] = s_{m(A,B)} \), so:

\[
s^*_{(A,B)} = \sum_{m=1}^M w_m s_{m(A,B)}.
\]

By this derivation, Gleason's theorem proves that the aggregated probability of ``Document A being more relevant than Document B" is the linear weighting of the individual pairwise probabilities from each base ranker. This validates the pairwise fusion model as theoretically optimal, mirroring the pointwise case but adapted for relative ordering in reranking tasks. The weights \( w_m \) can be learned interactively (as described in the main text), enabling efficient aggregation without full supervision.

\section{Prompt}\label{appendix: prompt}
We ask the LLM to pick the more relevant of exactly two passages using the following prompt (shown for the decoder-only judges, e.g.\ Llama-3.1-8B-Instruct; the encoder-decoder Flan-T5 judges use the same user turn without a system message), matching the pairwise, two-label (A/B) setup described in \S\ref{sec: methodology}; since the LLM is requested to output only a single character, the number of output tokens is negligible.

\begin{tcolorbox}
\textbf{System:} You are a helpful passage-ranking assistant. Read the query and the candidate passages carefully, then output ONLY the single letter label (A, B) of the passage most relevant to the query.

\textbf{User:} Given a query ``\{query\}'', which of the following passages is the most relevant to the query?

Passage A: ``\{passage\_A\}''

Passage B: ``\{passage\_B\}''

Output only the passage label of the most relevant passage:
\end{tcolorbox}

\end{document}